\documentstyle[sprocl]{article}

\input{psfig}

\begin{document}

\title{Mass Follows Light}

\author{C.S. Kochanek}

\address{Center for Astrophysics, MS-51 \\
60 Garden St., Cambridge, MA 02138, USA\\E-mail: ckochanek@cfa.harvard.edu} 


\maketitle\abstracts{ We use comparisons between the shapes of 
gravitational lens galaxies and models for their mass distributions
to derive statistical constraints on the alignment of the mass
distribution relative to the observed lens galaxy  and on the
strength of tidal shear perturbations.  The mass distributions are
aligned with the luminous galaxies, with a 
$\langle\Delta\theta^2\rangle^{1/2}<10^\circ$ upper limit on the
dispersion in the angle between the major axes.  
Statistical constraints, such as our bound on the misalignment 
between mass and light, are an important new approach to reducing 
the uncertainties in individual lens models, particularly for lenses 
used to estimate the Hubble constant.
}

\def\gtorder{\mathrel{\raise.3ex\hbox{$>$}\mkern-14mu
             \lower0.6ex\hbox{$\sim$}}}
\def\ltorder{\mathrel{\raise.3ex\hbox{$<$}\mkern-14mu
             \lower0.6ex\hbox{$\sim$}}}

\section{Introduction}

The shape of the dark matter halo surrounding a luminous galaxy is a powerful means
of exploring the mutual interactions of the two components.
Pure dark matter halos tend to be flattened and triaxial (e.g. Bardeen et al. 1986;
Frenk et al. 1988, Dubinski \& Carlberg 1991, Warren et al. 1992, also Bullock,
Moore, and Navarro in these proceedings), but the shape is then modified by interactions 
with the cooling baryons which reduce the triaxiality and the axis ratios  
(e.g. Evrard et al. 1994, Dubinski 1994). 
Many lines of observational evidence indicate that luminous galaxies are nearly oblate 
(see the review by Sackett 1999).  While the halos of disk galaxies are rounder than the stellar
distribution, Buote \& Canizares (1996, 1997, 1998) typically found that
the flattening of the X-ray isophotes of elliptical galaxies implied a mass distribution
flatter than the luminous galaxy.

The over 60 known gravitational lenses provide a newer laboratory for comparing
the mass and the light.  Both the projected mass enclosed by the Einstein ring and the
projected quadrupole of the gravitational field at the ring are accurately measured with
little dependence on the details of the model (see Kochanek 1991, Wambsganss \& Paczynski 1994,
Witt \& Mao 1997).  The lensed images are typically 1-1.5 effective radii from the
center of the lens galaxy where the quadrupole will have significant contributions from
both the luminous galaxy and the dark matter halo.  Even interior to the Einstein ring,
there is a significant contribution of the dark matter to the enclosed mass (about half)
and the matter inside and outside the Einstein ring make comparable contributions
to the overall quadrupole.  Thus, by comparing the gravitational field to the luminous lens galaxy
we can determine whether the mass and the light are aligned and whether they
have similar shapes.  Keeton et al. (1998) presented a preliminary
case for alignment but found little evidence for a correlation in the axis ratios.

On a practical level, modeling uncertainties for gravitational lenses could be
considerably reduced if the shape and orientation of the lens galaxy could be
used to constrain the mass distribution.  CMB anisotropy models have strong
degeneracies between many cosmological parameters and the Hubble constant
(e.g. Eisenstein, Hu \& Tegmark 1998), increasing the need for precise,
independent determinations of the Hubble constant.  It is unlikely that the
local distance ladder (e.g. Mould et al. 2000) will exceed 10\% accuracy, so
it is important to find new methods for measuring $H_0$.  The most promising
new approach is to use time delay measurements in gravitational lenses.
There are 9 lenses with time delay measurements (see Schechter 2000), although
many still have large uncertainties in the delays.  However, once the delay errors
are negligible, the only important  sources of uncertainty in the estimates
of the Hubble constant are the systematic errors in the lens models needed to
interpret the delays.  These systematic errors arise from the
paucity of constraints on the models of most lenses compared to the number of
parameters necessary for realistic mass models which include our uncertainties
about the structure of galaxies.  We can obtain more constraints either by
finding additional lensed structures, particularly Einstein ring images of
the source's host galaxy (see Kochanek et al. 2001), by using the
observed properties of the lens galaxy as constraints on the mass
distribution, or by using the lens sample as a whole to derive statistical
constraints on the mass distribution.

The radial distribution of the dark matter clearly differs from that of the
luminous lens galaxy, as constant mass-to-light ratio ($M/L$) models for
lenses generally work poorly (e.g. Lehar et al. 2000, 
Keeton et al. 2000, Cohn et al. 2001).  We may, however, be able to
use the shape of the luminosity distribution, the orientation of its major
axis and its axis ratio, as a constraint on the shape of the mass distribution.
We must, however, have clear experimental evidence for such a correlation.
In \S2 we examine the alignment of the mass with the light using
simple statistical methods. In \S3 we discuss statistical
lens constraints in general, and in \S4 we summarize our conclusions.

\section{The Alignment of Mass and Light}

We selected 20 lenses from the CfA-Arizona Space Telescope Lens Survey (CASTLES,
Falco et al. 2000) based on two criteria.  First, the HST
data had to be of sufficient quality to determine the ellipticity and orientation
of the primary lens galaxy. Second, there should be a single dominant 
lens galaxy.  The HST images were fit with photometric models including the lensed 
images and the lens galaxy following the procedures outlined in Lehar et al. (2000).  
We also modeled each lens as a singular isothermal ellipsoid (SIE) and determined 
the best fit model axis ratios, orientations and their uncertainties.

\begin{figure}[t]
\centerline{\psfig{figure=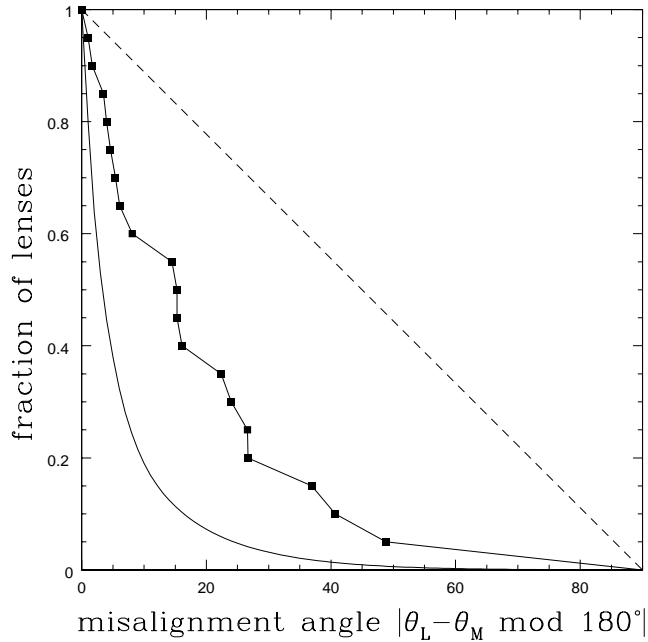,width=3.5in}}
\caption{ The misalignment angle distribution. The curve connecting the points 
  shows the distribution for the lens sample.  The solid (dashed) curve shows
  the expected distribution if the properties of the mass and the light are
  perfectly correlated (random).
  }
\end{figure}
\begin{figure}[t]
\centerline{\psfig{figure=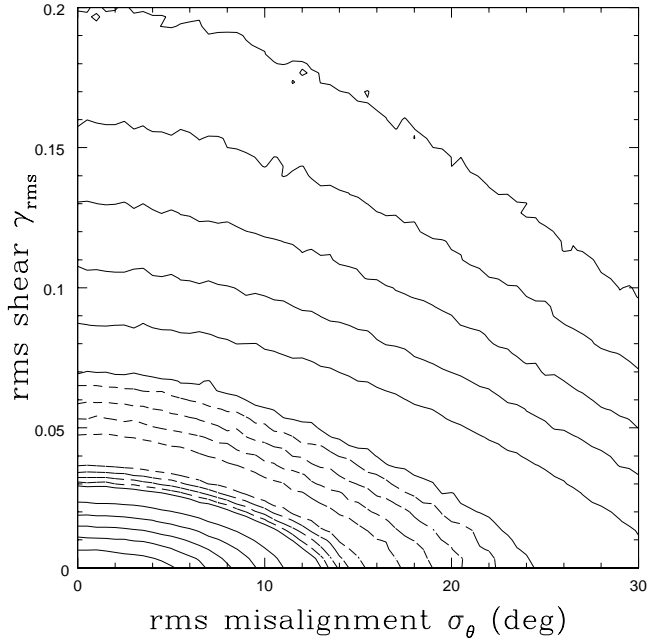,width=3.5in}}
\caption{
  Logarithmic contours of the Kolmogorov-Smirnov test probabilities for matching the observed
  misalignment angle distribution as a function of the rms misalignment $\sigma_\theta$ and the
  tidal shear $\gamma_{rms}$.  We expect tidal shears of $\gamma_{rms}\simeq 0.06$.
  The solid contours are logarithmically spaced by $0.5$~dex and the dashed contours are
  spaced by $0.1$~dex for log likelihoods between 0 and
  $-0.4$.  The differences between the dashed contours are not statistically significant.
  }
\end{figure}

\def\vecg{\vec{\gamma}}
\def\vece{\vec{\epsilon}}

We can measure the misalignment between the luminous lens galaxies and the SIE models
by the misalignment angle
$\Delta\theta = | (\theta_L -\theta_M) \, \hbox{mod} \,180^\circ |$ where $\theta_L$
and $\theta_M$ are the major axis position angles of the galaxy and the model
with $0^\circ \leq \Delta\theta \leq 90^\circ$.  If the two angles
were uncorrelated, then the integral distribution would be linear, with
$P(>\Delta\theta)=1-(\Delta\theta/90^\circ)$.  If the two angles were perfectly
correlated, we would still see a finite width distribution because of the
uncertainties in estimating both $\theta_L$ and $\theta_M$.  If the uncertainty in
measuring $\Delta\theta$
is $\sigma \ll 90^\circ$, then we would expect $P(>\Delta\theta) = \hbox{Erfc}(\Delta\theta/\sqrt{2}\sigma)$
where Erfc$(x)$ is the complementary error function.
Fig. 1 shows that the observed distribution of misalignment angles matches neither
the random nor the correlated distributions.
The Kolmogorov-Smirnov (K--S) test probabilities that the data are drawn from the random or
perfectly correlated distributions are $4\times 10^{-6}$ and $3\times 10^{-4}$
respectively.

Before we can draw any conclusions about misalignments between the mass and the light,
we must compensate for the effects of tidal shear perturbations.  All lenses are
modified by tidal shear perturbations due to nearby galaxies and the group or cluster
halo in which the lens resides (see Keeton et al. 1997), or perturbations from large scale
structure (see Barkana 1996, Keeton et al. 1997)
along the ray.  The lens model fits the average quadrupole of the gravitational
field, which is a combination of the quadrupole from the lens galaxy and the tidal
shear.  We can define the shapes of the light, the mass and the model and the tidal
shear by the pseudo-vectors
  $\vec{\epsilon}= \epsilon \lbrace \cos 2\theta_\epsilon, \sin 2\theta_\epsilon \rbrace$
and
  $\vec{\gamma}= \gamma \lbrace \cos 2\theta_\gamma, \sin 2\theta_\gamma \rbrace$
where $\epsilon$ is the ellipticity ($\epsilon = 1-b/a$ for axis ratio $0\leq b/a\leq 1$),
$\gamma$ is the magnitude of the shear, and
$\theta_\epsilon$ and $\theta_\gamma$ are the orientations of the major axes.  There
are four ``shapes'' associated with each lens $i$:
 the shape of the luminous galaxy, $\vece_{L i}$,
 the shape of the lens model, $\vece_{M i}$,
 the shape of the mass distribution of the lens galaxy, $\vece_{G i}$, and
 the external shear at the lens galaxy, $\vecg_i$.
A good approximation is that lens models which produce the same Einstein
ring shape will be nearly indistinguishable (see Keeton et al. 1997,
Kochanek et al. 2001), which corresponds to the constraint
  $\vece_{M i} = \vece_{G i} + 6 \vecg_i$.
The factor of $6$ is for SIE models, but has little effect on the conclusions.

\begin{figure}[t]
\centerline{\psfig{figure=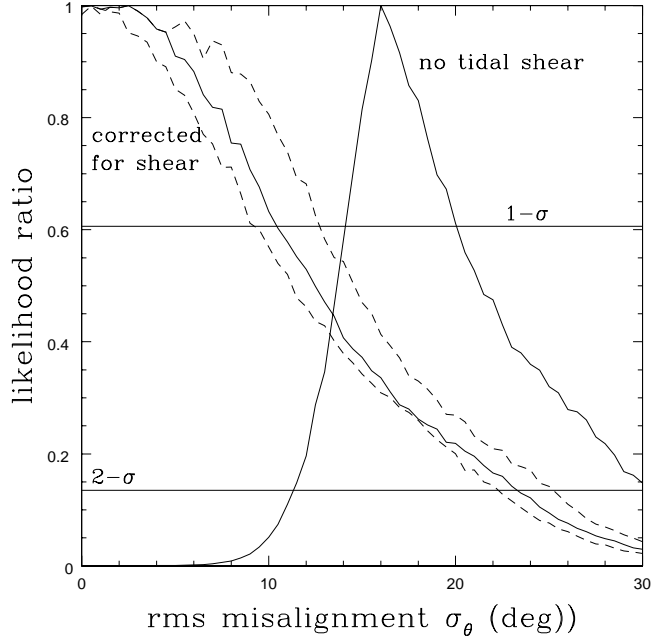,width=3.5in}}
\caption{Likelihood ratios for the dispersion of the intrinsic misalignment distribution
 $\sigma_\theta$.  The solid curve labeled ``no tidal shear'' shows that a
 $\sigma_\theta=16^\circ$ width is needed to explain the misalignment distribution in the
 absence of any tidal shear.  The solid line labeled ``corrected for tidal shear'' shows that
 the theoretically expected tidal shears are large enough to explain the misalignment
 distribution without any intrinsic misalignment.  The dashed curves show the effect of
 increasing and decreasing the prior estimate of $\gamma_{rms}$ by 50\%. }
\end{figure}

By studying the correlations between the shapes of the models, $\vece_M$, and the
shapes of the luminous lens galaxies, $\vece_L$, we can statistically estimate the
correlations between the shapes of the mass and the light and the strength of the
external tidal shears.
For our first analysis we will examine only the misalignment angle distribution.  We
assume that $\epsilon_G \simeq \epsilon_M$ as an estimate for the ease with which the
external shear can twist the major axis of the model away from the major axis of the
mass distribution.  If $\theta_G-\theta_L$ is the misalignment between
the mass and the light, the simplest model for its distribution is a
Gaussian of width $\sigma_\theta$,
\begin{equation}
 P(\theta_G-\theta_L) =
    \exp(-(\theta_G-\theta_L)^2/2\sigma_\theta^2)/\sqrt{2\pi} \sigma_\theta.
\end{equation}
We wrap the tails of the Gaussian onto the $0^\circ$ to $90^\circ$ range of the 
misalignment angle.  We assume the two shear components are Gaussian variables, so the
probability distribution for the total shear is
\begin{equation}
 P(\gamma) =  2 \gamma \exp(-\gamma^2/\gamma_{rms}^2)/\gamma_{rms}^2
\end{equation}
where $\gamma_{rms}$ is the rms shear.  The shape of the shear distribution is not
critical because of the small sample size.  We used Monte Carlo simulations to model 
the misalignment angle distribution including the effects measurement errors, and then 
used the K--S test to see if the
distribution agreed with the data.  Fig. 2 shows contours of the K--S test probability
as a function of the rms misalignment angle, $\sigma_\theta$, and the strength
of the tidal shear, $\gamma_{rms}$.
Although the sample is too small to determine the origin of the misalignments,
we are restricted to a narrow parameter range.  In the limit where there is no tidal
shear, the width of the misalignment distribution must be $\sigma_\theta = 16^\circ$
($14^\circ \ltorder \sigma_\theta \ltorder 20^\circ$), while in the limit where there
are no misalignments, the rms shear must be $\gamma_{rms} = 0.04$ 
($0.033 \ltorder \gamma_{rms} \ltorder 0.055$), where we have
used maximum likelihood methods to convert the K-S probability distributions into
uncertainties.

We can measure the dispersion $\sigma_\theta$ by adding a prior probability for the strength 
of the tidal shear based on the survey of shear in lenses by Keeton et al. (1997).  The typical 
perturbation from large scale structure is $\gamma_{rms} \simeq 0.03$ based on standard
normalizations for the non-linear power-spectrum (see Barkana 1996, Keeton et al. 1997),
and the typical perturbation from objects clustered with the lens produces an effective
perturbation of $\gamma_{rms}\simeq 0.05$ (see Keeton et al. 1997).  The total rms shear 
is then $\gamma_{rms} \simeq 0.06$ after adding the two terms in quadrature.
We include this estimate as a log-normal prior for $\gamma_{rms}$, assuming that its 
value is uncertain by a factor of two.  After adding the prior to the likelihoods, 
the best models have $\sigma_\theta\simeq 0^\circ$ and an upper bound of 
$\sigma_\theta < 10^\circ$ on the rms misalignment.  The results are not sensitive to the 
exact value we use for the prior, as raising $\gamma_{rms}$ by 50\% lowers the bound to 
$9^\circ$ and lowering it by 50\% raises the bound to $12^\circ$.  Fig. 3 shows the 
likelihood ratios as a function of the rms misalignment angle.  

\section{Statistical Lens Constraints}

Our results on the alignment of the mass and light are an example of a broader
approach to improving the constraints on gravitational lenses.  The problem with 
models of individual lenses is that they frequently have too few constraints to 
avoid degeneracies between physically interesting parameters.  This is 
particularly vexing for the gravitational lenses with time delays, where there
is usually a degeneracy between the value of $H_0$ and the radial density profile
of the lens galaxy (see, e.g., Impey et al. 1998, Witt et al. 2000).  Most of
these problems can be eliminated using statistical constraints from the properties of 
the lens population.

One example of a statistical constraint method is the bounding of the central densities
of lenses, most recently by Rusin \& Ma (2001).  Almost all lenses lack
visible central (``odd'') images, which sets a lower bound on the central density 
of a successful model.  As an example, consider a modern density model
with a central density cusp $\rho \sim r^{-\beta}$ ($1 \ltorder \beta \ltorder 2$) inside break 
radius $a$, and a steeper $\rho \sim r^{-n}$ ($3 \ltorder n \ltorder 4$) profile outside.
The models of most lenses will have a degeneracy between the physical parameters
$a$ and $\beta$ in which the break radius decreases as the central cusp is softened
so as to keep the central density high enough to make the central image invisible
(see Munoz et al. 2001).  The ellipticity of the lens also rises as the mass becomes
more centrally concentrated in order to keep the same quadrupole moment at the Einstein 
ring.   However, most of these permitted models for the individual
lenses imply statistical properties for the lens population which are not consistent
with the observed properties.  Rusin \& Ma (2001) used this
to show that for models with break radii larger than the Einstein ring radius, only
cusps steeper than $\beta \gtorder 1.8$ were statistically consistent with the data.

Our limit on the alignment of mass and light is a second example of a statistical
constraint.  We know that successful models of individual lenses require contributions 
from both the ellipticity of the lens and external tidal perturbations but cannot determine 
all their parameters without degeneracies (see Keeton et al. 1997).  However, by performing 
a joint analysis of the models for a sample of lenses, we have shown that the mass and 
luminosity distributions of the lens are aligned within a reasonably tight statistical 
limit of $\langle \Delta \theta^2\rangle^{1/2} < 10^\circ$.  We can also assign a 
prior on the typical strength of the tidal shear with $\gamma_{rms}\simeq 0.05$.  These
two restrictions can now be used to reduce the parameter degeneracies in lens models.  

The next step from these two simple examples is to attack the problem of statistical
constraints formally.  There are two complimentary approaches. First, models of 
individual lenses should be weighted by the likelihood of observing them given 
the model parameters.  This is the generalization of the limits on central densities.
Second, the statistical properties of lens samples should be homogeneous.  This
is the generalization of our bound on the alignment of mass and light.

In our first generalization we consider models of a particular lens, where we can
add the statistical constraint that models in which the observed configuration is 
probable should be favored over those where it is not. Consider the case of a density 
cusp with parameters $\beta$ and $a$ where in fitting the observed properties of the 
lens the fit statistic $\chi^2(\beta,a)$ has a degeneracy between the parameters.  For 
any model ($\beta$, $a$) we can also calculate the probability $p_{tot}(\beta,a)$ that 
a lens with such physical properties will be found, and the probability $p_{obs}(\beta,a)$
that it will have the observed configuration.  If two models are consistent with the
observed configuration, we should statistically favor the model in which the observation
is more likely by adopting the modified likelihood
\begin{equation}
 2 \ln L(\beta,a) = -\chi^2(\beta,a) \rightarrow 
    -\chi^2(\beta,a) + 2 \ln (p_{obs}(\beta,a)/p_{tot}(\beta,a)).
\end{equation}
While for a particular lens this may bias the solution, for an ensemble of lenses
it should be correct.  The Rusin \& Ma (2001) limit on the cusp exponent can be
thought of as a variant of this general approach. 

In our second generalization we consider (Bayesian) models for a population of lenses. 
For the problem of 
ellipticity and shear we considered in \S2, the fit statistic $\chi^2_i(\vec{p}_i)$
for lens $i$ depends on the four parameters for the angular structure, 
$\vec{p_i}=(\epsilon_{Gi}, \theta_{Gi}, \gamma_i, \theta_{\gamma i})$. Our
calculation in \S2 simplified the $\chi^2$ by assuming it could be
approximated by a degenerate constraint, $\vec{\epsilon}_M=\vec{\epsilon}_G+6\vec{\gamma}$,
on the variables. In a Bayesian formalism, the probability of the data $D_i$ given
the parameters $\vec{p}_i$ is $P(D_i|\vec{p}_i) \propto \exp(\chi_i^2(\vec{p})_i/2)$,
and we use Bayes theorem ($P(p|D)\propto P(D|p)P(p)$) to estimate the probability of the
parameters given the data and the priors.  Several of the parameters have non-uniform
Bayesian priors.  In particular, the major axis of the mass is constrained by that
of the light by the prior $P(\theta_{G_i}|\theta_{Li},\sigma_\theta)$ given in eqn. (1),
and the shear is constrained by the rms shear by the prior $P(\gamma_i|\gamma_{rms})$ 
given in eqn. (2).  These priors introduce new parameters, $\gamma_{rms}$ and 
$\sigma_\theta$, which require their own priors, $P(\gamma_{rms})$ and 
$P(\sigma_\theta)$.  Thus, the Bayesian probability for a sample $i=1\cdots N$ lenses is
\begin{equation}
P(\vec{p}_i,\gamma_{rms},\sigma_\theta|D_i) \propto P(\gamma_{rms})P(\sigma_\theta)
  \Pi_{i=1}^N P(\gamma_i|\gamma_{rms}) P(\theta_{G_i}|\theta_{Li},\sigma_\theta)
   P(D_i|\vec{p}_i).
\end{equation}
While this looks formidable, we can find the probability distribution for any particular
variable simply by marginalizing (integrating) over the remaining variables.  For example,
the Bayesian equivalent of the likelihood shown in Fig. 2 is found by integrating over
all the lens parameters,
\begin{equation}
   P(\gamma_{rms},\sigma_\theta|D_i) \propto 
      P(\gamma_{rms})P(\sigma_\theta) 
    \Pi_{i=1}^N \int d \vec{p}_i P(\gamma_i|\gamma_{rms}) 
       P(\theta_{G_i}|\theta_{Li},\sigma_\theta) P(D_i|\vec{p}_i).
\end{equation}
For variables constrained by the data, the final 
probability distribution will be significantly narrower than the prior distribution 
($P(\gamma_{rms})P(\sigma_\theta)$).  Our analysis in \S2 is a simpler, more intuitive
implementation of this approach, but it is easier to add parameters or to
include terms like those from eqn. (3) in a more general Bayesian formalism.

\section{Conclusions}

Almost all the lens galaxies we consider are massive early-type galaxies, which have
the most complex formation history of normal galaxies.  First, CDM prolate-triaxial 
halos form in which the baryons cool into a disk and form stars.  The cooling baryons 
compress the dark matter and the oblate shape of the baryonic potential begins to 
destroy the triaxiality of the dark matter.  After the halo undergoes several major mergers, 
the disks are destroyed to leave a nearly oblate early-type galaxy with some fossil 
evidence for the disks.  Despite this complicated formation history we can show that
the final mass and luminosity distributions are aligned to within 
$\langle \Delta\theta^2 \rangle^{1/2} < 10^\circ$ (in projection) by examining
the statistics of the misalignments between lens models and lens galaxies.   The
quadrupole of the mass includes similar contributions from both the luminous and
the dark matter, so a small misalignment between the mass and the light should
imply a small misalignment between the dark and the light.

Our approach to measuring the misalignment of mass and light is an example of a
statistical lens constraint.  Where models of individual lenses frequently have
awkward degeneracies between physical parameters, the most famous of which is 
that between the Hubble constant and the radial mass distribution of the lens,
samples of lenses generally should not.  To lay the basis for this new approach
to lens modeling, we have outlined two general and complimentary approaches to
developing statistical constraints for gravitational lenses, one of which is
a generalization of the methods we used to study the alignment of mass and light. 
Given a sample of lenses with time delay measurements (there are now 9 
lenses with time delays), these statistical approaches should greatly reduce 
the uncertainties in estimates of the Hubble constant over that in any single 
system.

\noindent {\bf Acknowledgments:} CSK thanks the remainder of the CASTLES 
collaboration for its comments. Support for the CASTLES project was provided 
by NASA through grants GO-7495, GO-7887, GO-8175 and GO-8804 from STScI.  
CSK is also supported by the Smithsonian Institution and NASA grants 
NAG5-8831 and NAG5-9265.

\end{document}